\documentclass[aps,prx,twocolumn,showpacs,superscriptaddress,email]{revtex4-1}
\usepackage[latin1]{inputenc}
\usepackage{amsmath} 
\usepackage{graphicx} 
\usepackage{float}
\usepackage{dcolumn}   
\usepackage{bm}        
\usepackage{amssymb}   
\usepackage[latin1]{inputenc}

\usepackage{xcolor}
\begin{document}

\title{\bfseries{Hidden symmetries in plasmonic gratings}}
\author{P. A. Huidobro}
\email[p.arroyo-huidobro]{@imperial.ac.uk}
\affiliation{Imperial College London, Department of Physics, The Blackett Laboratory, London SW7 2AZ, UK}
\author{Y.H. Chang}
\affiliation{Imperial College London, Department of Physics, The Blackett Laboratory, London SW7 2AZ, UK}
\author{M. Kraft}
\affiliation{Imperial College London, Department of Physics, The Blackett Laboratory, London SW7 2AZ, UK}
\author{J. B. Pendry}
\affiliation{Imperial College London, Department of Physics, The Blackett Laboratory, London SW7 2AZ, UK}

\date{\today}

\begin{abstract}
Plasmonic gratings constitute a paradigmatic instance of the wide range of applications enabled by plasmonics. While subwavelength metal gratings find applications in optical biosensing and photovoltaics, atomically thin gratings achieved by periodically doping a graphene monolayer perform as metasurfaces for the control of terahertz radiation. In this paper we show how these two instances of plasmonic gratings inherit their spectral properties from an underlying slab with translational symmetry. We develop an analytical formalism to accurately derive the mode spectrum of the gratings that provides a great physical insight. 
\end{abstract}

\maketitle

\section{Introduction}
Plasmonic systems display a cornucopia of electromagnetic phenomena primarily dictated by surface geometry. In previous papers we have striven to bring some order to the many diverse phenomena by showing that systems can have identical spectra even though they may have very different geometry \cite{Pendry2012,Luo2014}. In this paper we apply our theory to extended systems, specifically to a class of gratings related by conformal transformations. In particular we show that a whole class of gratings share their properties with a flat surface and as a result have surprising spectral properties where the regular discrete translational symmetry of the grating is superseded by the continuous symmetry of a flat surface. This symmetry is only broken for some of the modes away from normal incidence which exhibit a band gap, and this paper deals with this.

Our tool of choice is transformation optics \cite{Ward1996,Pendry2006} which shows how the parameters $\epsilon$ and $\mu$ in Maxwell's equations change when one geometry is transformed into another. For example when a cylinder is transformed into an ellipse \cite{Kraft2014}, or when a knife edge is transformed into a series of waveguides \cite{Luo2010a}. The tool is particularly powerful in two dimensional systems where conformal transformations leave the in-plane components of $\epsilon$ and $\mu$ unchanged. Furthermore in plasmonic systems the dimensions are often measured in nanometres so that surface plasmons which are largely electrical in character have very small out of plane magnetic fields that can be neglected. Here we can concentrate on the in-plane response and when needed treat magnetic components as a perturbation. Out of plane $\epsilon$ and $\mu$ are changed by the transformations and in particular the magnetic fields always see the periodicity of the grating through variations in $\mu$.

\begin{figure}[p]
\begin{center}
\includegraphics [angle=0, width=0.9\columnwidth]{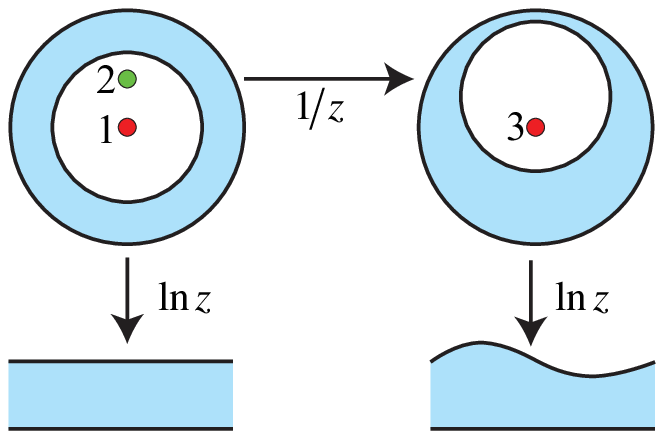}
\caption[]{Two dimensional Plasmonic structures related by conformal transformations. The finite cross section annular cylinder at top left is related to the infinite structures in the lower half of the figure which inherit its spectral properties, but with a twist given by singularities in the transformations.}\label{figdrawing}
\end{center}
\end{figure}

In a previous paper \cite{Kraft2015} we investigated the properties of plasmonic gratings derived from conformal transformations and found analytic expressions for the grating's response to light incident normal to the grating surface. In that paper we noted the problems presented by non normal incidence but left them unresolved. Here we return to the problem and present a complete solution. In Fig. 1 we sketch the structure of the transformations. Starting from a metal cylindrical annulus we can either make a $\ln z$ transformation, using point 1 as the origin, directly to a continuous slab of metal, where the angle around the cylinder maps into the horizontal direction and the radial direction into the vertical direction. Alternatively we can invert about point 2 to give the crescent on the right. A further transformation of $\ln z$, using point 3 as the origin, the centre of the larger circle, leads to a continuous grating structure of finite thickness shown on the lower right. When the grating is not too strongly modulated, the shape of the surface is well approximated by a sine wave. By varying the parameters a whole class of gratings can be generated. We can also apply our theory to periodically doped graphene by taking the limit of a very thin grating \cite{Huidobro2016,Huidobro2016a}.

In this way a flat slab of metal is conformally related to many gratings that inherit the spectral properties of an annular cylinder and of a flat surface. Except that there is a catch. Previous work related infinite structures to finite structures which contained a geometric singularity and had the strange property for a finite system of inheriting the continuous spectrum of the infinite parent. Here our finite structure with a discrete spectrum is related to an infinite structure with a continuous spectrum. The requirement that the plasmonic solutions in the cylinder are continuous at all angles quantises the spectrum. However the infinite slab and grating have additional modes that are not directly related through the transformation. Again a singularity is the culprit and the transformation illustrated in Fig. 2 implies a branch cut. The Bloch wave solutions in the infinite cases only map onto the finite solutions of the cylinders at the zone centre where solutions have the same periodicity as the lattice. The challenge addressed in this paper is to build into our theory solutions away from the zone centre. This is important for a complete understanding of these systems as non-central wave vectors comprise the great majority of solutions. 

Previously we showed that at the centre of the Brillouin zone, $k=0$, the grating structure effectively disappears: there are no band gaps and mode frequencies are the same as for the slab, and the group velocity remains finite. At $k=0$ external waves incident on the grating transform into waves incident on the slab: the properties of the grating are predicted by the response of the slab. This result is only slightly disturbed by inclusion of magnetic perturbations. Away from $k=0$ this result ceases to hold and, as well show below, so distorting is the geometric singularity that some of the incident wave has to be reclassified as though it were in fact a reflected wave. This paper shows how to take account of these distortions and make accurate calculations of the grating spectrum.

This manuscript is organised as follows. In section \ref{sec:methodology} we detail our analytical formalism to study the band structure of subwavelength plasmonic gratings. Next, in section \ref{sec:results} we apply our method to study two instances of plasmonic gratings. First, we consider a graphene monolayer subject to a periodic bias potential, where the periodic modulation of the conductivity acts as a subwavelength grating. Second, we study periodic gratings in thin metal slabs. 

\section{Methodology}
\label{sec:methodology}
\begin{figure*}[p]
\begin{center}
\includegraphics [angle=0, width=0.9\textwidth]{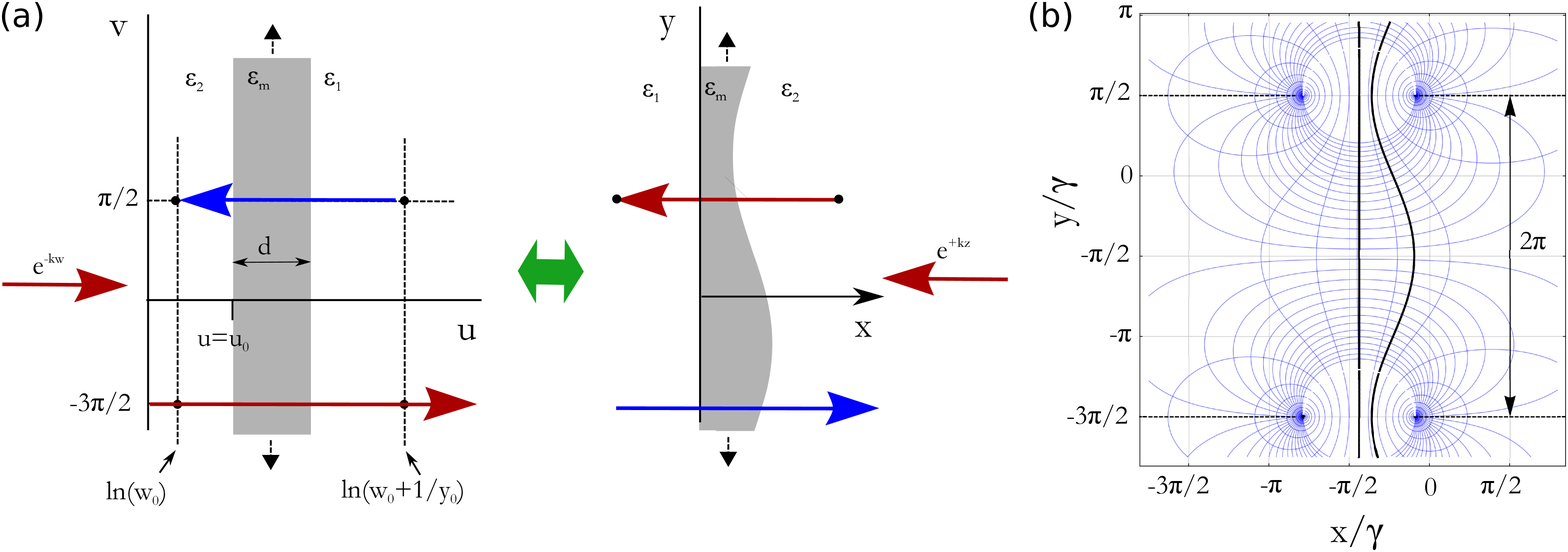}
\caption[]{(a) A periodic metal grating (right) maps to an infinite slab through a conformal transformation. The transformation results in the mixing of some waves between the slab (left) and grating (right) frames. The black circles denote singular points at $w_1=\ln(w_0)+i\pi/2$ and $w_1=\ln(w_0+y_0^{-1})+i\pi/2$ (with $2\pi$ periodicity along the vertical direction). (b) Contour lines for the conformal transformation that relates the infinite slab to the plasmonic grating, Eq. \ref{eq:transformation}. Lines of constant $u$ and $v$ are plotted. The parameters used are: $\gamma=1$, $w_0=1.5$, $u_0=1$ and $d=0.5$. The thick black lines correspond to the maps of $u=u_0$ (wavy line) and $u=u_0+d$ (straight line), and represent the grating shape.} \label{figsketch}
\end{center}
\end{figure*}

In this section we show how we can relate subwavelength plasmonic gratings to flat slabs using transformation optics \cite{Kraft2015,Huidobro2016,Huidobro2016a}. We start from a simple flat slab of plasmonic material of thickness $d$ placed at $u=u_0$ [see Fig. \ref{figsketch} (a), left]. The permittivity of the slab is $\epsilon_m(\omega)$, and of the surrounding dielectrics is $\epsilon_{1,2}$. Our methodology can treat metal gratings \cite{Kraft2015} or, in the limit of an infinitesimally thin slab, a graphene sheet with periodically modulated doping \cite{Huidobro2016,Huidobro2016a}. Here we develop our formalism for a general $d$ and $\epsilon_m$, and in section \ref{sec:results} we study the two cases in detail.

Making use of the conformal transformation \cite{Kraft2015,Huidobro2016,Huidobro2016a}
\begin{equation}
	z=\gamma\ln\left(  \frac{1}{e^w -iw_0} + i y_0 \right)
	\label{eq:transformation}
\end{equation}
the infinite slab in the left hand side of Fig. \ref{figsketch} (a) can be mapped to the periodic grating shown in the right hand side of that figure. In Eq. \ref{eq:transformation}, $w=u+iv$ refers to the coordinates in the slab frame, and $z=x+iy$ to the coordinates in the grating frame. Figure \ref{figsketch}(b) shows a plot of the conformal transformation given by Eq. \ref{eq:transformation} for the values of $\gamma$, $w_0$ and $y_0$ given in the figure caption. The branch points (which repeat periodically along the $y$ axis with period $2\pi \gamma$) and branch cuts (which extend from the branch points out to infinity) modulate the shape of the contour lines. This results in wavy contours that repeat themselves along the $y$ direction and that can be used to model a grating by working in the region enclosed by the two sets of branch points, as depicted with a thick black line. In Eq. \ref{eq:transformation}, $\gamma$ is an overall scale factor that determines the size of the structure through the period of the grating, $a=2\pi\gamma$. In addition, $w_0$ is a free parameter that sets the grating modulation strength, and $y_0$ is fixed by $w_0$, the slab thickness, $d$, and its position, $u_0$, as $y_0=w_0/ \left[ \exp\left(2(u_0+d)\right) -w_0^2 \right]$. The position of the slab is chosen so that one of its boundaries transforms to a straight line in the $z$ plane such that only one of the grating sides is modulated (see Fig. \ref{figsketch}).  

In the electrostatic limit, the spectral properties of a plasmonic structure are only determined by geometry. Subwavelength plasmonic gratings, of period $a\ll\lambda$, with $\lambda=2\pi/k_0$ being the free space wavelength, can be well described within that limit. Recalling that conformal transformations applied to Maxwell's equations conserve not only the in-plane material parameters, but also the electrostatic potential \cite{Pendry2002}, we conclude that the slab and grating frames are equivalent within the electrostatic limit. Therefore finding the modes of the more simple translationally invariant slab automatically gives the modes of the grating, which has a more complicated shape. This means that the band structure of the grating can in principle be obtained by simply taking the dispersion relation of the plasmonic slab in the electrostatic limit and folding it into the first Brillouin zone \cite{Kraft2015}. 

However, in order to appropriately calculate the modes in the grating frame, one must take into account the boundary conditions to be imposed in the slab frame. Since the system in the grating frame is periodic along $y$, one needs to impose the same periodicity in the $(u,v)$ frame, where the dispersion relation is continuous. Nevertheless, solutions at a finite wave vector $k$ have an unphysical discontinuity of the phase across the branch cuts of Eq. \ref{eq:transformation}. For this reason, the equivalence between the translationally invariant slab and the periodic grating strictly holds only close to the zone centre, for $k\ll\pi/a$ \cite{Kraft2015}. In the following we develop a method that allows us to accurately predict the modes of the grating even at the zone boundary, $k=\pi/a$. 

Our procedure consists of studying the response of the grating to incident waves {\it decaying} towards the grating. The causal response will consist of waves {\it growing} toward the grating. We start by considering an incident wave from the right onto the grating, with wavenumber $k_0=\sqrt{k_x^2+k_y^2}$. Within the electrostatic limit ($k=k_y\gg k_0$, $k_x\approx i k$ ) we write the electrostatic potential in different regions of space as follows, 
\begin{eqnarray}
\Psi^{x>0}_{inc}(x,y)  & = & \label{eq:Pinc1} \\
& = & \sum_{g}a_{k+g}^{+}\exp\left[+(k+g)z\right], \,  k+g>0, \nonumber \\
 & = & \sum_{g}a_{k+g}^{+}\exp\left[-(k+g)z^{*}\right], \, k+g<0, \nonumber \\
 \Psi^{x<0}_{inc}(x,y) & = & \label{eq:Pinc2} \\
 & = & \sum_{g}a_{k+g}^{-}\exp\left[-(k+g)z^{*}\right],\,  k+g>0, \nonumber \\
 & = & \sum_{g}a_{k+g}^{-}\exp\left[+(k+g)z\right],\,  k+g<0.  \nonumber 
\end{eqnarray}
Here, we sum over the modes of the grating, $g$, and $a_{k+g}^{\pm}$ are expansion coefficients of the incident waves. We note that we work with dimensionless coordinates and lattice vectors, this is, $k$ and $z$ stand for $k\gamma $ and $z/\gamma$, respectively. 
Similarly, for the reflected waves, 
\begin{eqnarray}
\Psi^{x>0}_{ref}(x,y)  & = & \\
& =&\sum_{g}b_{k+g}^{+}\exp\left[-(k+g)z^{*}\right], \,  k+g>0, \nonumber \\
 & =&\sum_{g}b_{k+g}^{+}\exp\left[+(k+g)z\right], \,  k+g<0, \nonumber \\
\Psi^{x<0}_{ref}(x,y) & = & \\
& =&\sum_{g}b_{k+g}^{-}\exp\left[+(k+g)z\right], \,  k+g>0, \nonumber \\
 & =&\sum_{g}b_{k+g}^{-}\exp\left[-(k+g)z^{*}\right], \,  k+g<0, \nonumber 
\end{eqnarray}
where $b_{k+g}^{\pm}$ stand for the expansion coefficients of the reflected waves. 

We wish to express the waves in the slab frame, where the boundary conditions are easier to apply. Since the electrostatic potential is conserved, one simply needs to replace $z$ by $w$ through their relation, Eq. \ref{eq:transformation}. As sketched in Fig. \ref{figsketch}(a), in transforming from the grating to the slab frame, waves go to opposite sides of the slab, from $x>0$ to $u<0$ and vice versa. Using Eq. \ref{eq:transformation}, we have for the incident waves, Eqs. \ref{eq:Pinc1} and \ref{eq:Pinc2},  
\begin{eqnarray}
 & & \Psi^{u<0}_{inc}(u,v)  =  \\
& =  & \sum_{g}a_{k+g}^{+}\left[\frac{1}{\exp(w)-iw_{0}}+iy_{0}\right]^{+(k+g)},  \,  k+g>0, \nonumber \\
 & = & \sum_{g}a_{k+g}^{+}\left[\frac{1}{\exp(w^{*})+iw_{0}}-iy_{0}\right]^{-(k+g)},  \,  k+g<0, \nonumber 
 \end{eqnarray}
 
\begin{eqnarray}
&&\Psi^{u>0}_{inc}(u,v) = \\
 & =&\sum_{g}a_{k+g}^{-}\left[\frac{1}{\exp(w^{*})+iw_{0}}-iy_{0}\right]^{-(k+g)},  \,  k+g>0, \nonumber \\
 & =&\sum_{g}a_{k+g}^{-}\left[\frac{1}{\exp(w)-iw_{0}}+iy_{0}\right]^{+(k+g)},  \,  k+g<0. \nonumber 
\end{eqnarray}
and similarly for reflected waves. 

By expanding Eq. \ref{eq:transformation} as a series, we can rewrite the potential for the incident waves in the slab frame in series form, 
\begin{widetext}
\begin{eqnarray} 
\psi_{inc}^{u<0}(u,v) &=& \\ 
 &=& \sum_{g}a_{k+g}^{+}\left\{ I_{g}e^{-w\left(k+g\right)}+\sum_{n=1}^{\infty}\left[Q_{n}^{+}\left(g\right)w_{0}^{n}e^{-\left(k+g+n\right)w}+Q_{n}^{-}\left(g\right)y_{0}^{\prime n}e^{-\left(k+g-n\right)w}\right]\right\},\, k+g>0 \nonumber \\
&=& \sum_{g}a_{k+g}^{+}\left\{ I_{g}e^{w^{\ast}\left(k+g\right)}+\sum_{n=1}^{\infty}\left[Q_{n}^{-}\left(g\right)w_{0}^{n}e^{+\left(k+g-n\right)w^{\ast}}+Q_{n}^{+}\left(g\right)y_{0}^{\prime n}e^{+\left(k+g+n\right)w^{\ast}}\right]\right\} ,\, k+g<0 \nonumber \\
\psi_{inc}^{u>0}(u,v)&=& \\ 
&=& \sum_{g}a_{k+g}^{-}\left\{ I_{g}e^{w^{\ast}\left(k+g\right)}+\sum_{n=1}^{\infty}\left[Q_{n}^{-}\left(g\right)w_{0}^{n}e^{+\left(k+g-n\right)w^{\ast}}+Q_{n}^{+}\left(g\right)y_{0}^{\prime n}e^{+\left(k+g+n\right)w^{\ast}}\right]\right\} ,\, k+g>0 \nonumber \\
&=&\sum_{g}a_{k+g}^{-}\left\{ I_{g}e^{-w\left(k+g\right)}+\sum_{n=1}^{\infty}\left[Q_{n}^{+}\left(g\right)w_{0}^{n}e^{-\left(k+g+n\right)w}+Q_{n}^{-}\left(g\right)y_{0}^{\prime n}e^{-\left(k+g-n\right)w}\right]\right\} ,\, k+g<0 \nonumber
\end{eqnarray}
\end{widetext}
where $y'_0 = y_0/(1+w_0y_0)$ and $I_g$ and $Q^{\pm}_n(g)$ are given in the Supplemental Material. The above series converges in the region between the two branch points, $\log|w_0|<u<\log|w_0+1/y_0|$. 
From these expressions we can see that, depending on the value of the exponents $k+g\pm n$, some of the waves change category:  from incident in the grating frame (see Eqs. \ref{eq:Pinc1} and \ref{eq:Pinc2}), they swap to reflected in the slab frame. Proceeding in a similar fashion for the reflected waves, one arrives at equivalent expressions where some of the waves change from reflected to incident (see Supplemental Material Eqs. S18, S19). The reason for this mixing of the waves is the change in boundary conditions alluded to above, since in the slab frame the waves are emitted from and disappear into singularities at, $w=w_1$ and $w=\log(e^{w_1}-e^{-y_1})$. 

In order to account for this, we swap some of the terms, such that all the incident and reflected waves are grouped in the same category in the slab frame. This gives, 
\begin{widetext}
\begin{eqnarray} 
 \psi_{inc}^{u<0}(u,v) &=& \\
&=&\sum_{g}\left\{ a_{k+g}^{+}\left[I_{g}e^{-\left(k+g\right)w}+\sum_{n=1}^{\infty}Q_{n}^{+}
\left(g\right)w_{0}^{n}e^{-\left(k+g+n\right)w}+
\sum_{n=1}^{g}Q_{n}^{-}\left(g\right)y_{0}^{\prime n}e^{-\left(k+g-n\right)w}\right] \right. +\nonumber \\ 
&&+ \left. b_{k+g}^{+}\sum_{n=g+1}^{\infty}Q_{n}^{-}\left(g\right)w_{0}^{n}e^{+\left(k+g-n\right)w^{\ast}}\right\},\, k+g>0 \nonumber \\
&=& \sum_{g}\left\{ a_{k+g}^{+}\left[I_{g}e^{\left(k+g\right)w^{\ast}}+\sum_{n=1}^{\infty}
Q_{n}^{-}\left(g\right)w_{0}^{n}e^{+\left(k+g-n\right)w^{\ast}}+\sum_{n=1}^{\left|g\right|-1}\left(Q_{n}^{+}\left(g\right)y_{0}^{\prime n}e^{+\left(k+g+n\right)w^{\ast}}\right)\right] \right.+  \nonumber \\ 
&&+ \left.b_{k+g}^{+}\sum_{n=\left|g\right|}^{\infty}Q_{n}^{+}\left(g\right)w_{0}^{n}e^{-\left(k+g+n\right)w}
\right\},\, k+g<0 \nonumber
\end{eqnarray}
\end{widetext}
where we have that the incident waves in the slab frame come from a combination of the incident and reflected waves in the grating frame, which are associated with coefficients $a_{k+g}^{\pm}$ and $b_{k+g}^{\pm}$, respectively.

Finally, the modes of the plasmonic grating can be derived by matching the fields at the slab interfaces. Introducing scattering matrices $\hat{\mathbf{T}}$ to account for the slab position at $u=u_0$, the matching equations read as, 
\begin{eqnarray}
\Psi_{ref}^{u>0} & = & {\mathbf{T}}^{++}\Psi_{inc}^{u<0}+{\mathbf{T}}^{+-}\Psi_{inc}^{u>0}\\
\Psi_{ref}^{u<0} & = & {\mathbf{T}}^{-+}\Psi_{inc}^{u<0}+{\mathbf{T}}^{--}\Psi_{inc}^{u>0}.
\end{eqnarray}
The scattering matrices, $\hat{\mathbf{T}}$, depend on the slab thickness, $d$, its position, $u_0$, and the plasmon dispersion relation in the slab frame (see Supplemental Material Section IV). Writing out this equation in terms of the expansion coefficients of the incident and reflected waves we have the following matrix equation, 
\begin{eqnarray}
\hat{\mathbf{S}} \hat{\mathbf{b}} + \hat{\mathbf{R}} \hat{\mathbf{a}} = \hat{\mathbf{T}} \hat{\mathbf{S}} \hat{\mathbf{a}} + \hat{\mathbf{T}} \hat{\mathbf{R}} \hat{\mathbf{b}} \label{eq:matrixeq}
\end{eqnarray}
where $\hat{\mathbf{a}} = (\mathbf{a}^+,\mathbf{a}^-)^T$, $\hat{\mathbf{b}} = (\mathbf{b}^-,\mathbf{b}^+)^T$, and,
\begin{eqnarray}
	\hat{\mathbf{T}} &=& \left(\begin{array}{cc} \mathbf{T}^{++} & \mathbf{T}^{+-} \\ \mathbf{T}^{-+} & \mathbf{T}^{--}  \end{array}\right),\, 
\end{eqnarray}	
\begin{eqnarray}
	\hat{\mathbf{S}} &=& \left(\begin{array}{cc} \mathbf{S}' & 0 \\  0 & \mathbf{S} \end{array}\right), 
\end{eqnarray}	
\begin{eqnarray}
	\hat{\mathbf{R}} &=& \left(\begin{array}{cc} 0 & \mathbf{R'} \\ \mathbf{R}  &0  \end{array}\right),
\end{eqnarray}
where the matrices $\mathbf{S}$, $\mathbf{S}'$, $\mathbf{R}$ and $\mathbf{R'}$ are obtained from the matching equation and are given in the Supplemental Material. Finally, in order to look for the eigensolutions of the grating we rearrange Eq. \ref{eq:matrixeq}, 
\begin{eqnarray}
 \left(\hat{\mathbf{R}} -  \hat{\mathbf{T}} \hat{\mathbf{S}} \right)  \hat{\mathbf{a}}  = \left(  \hat{\mathbf{T}} \hat{\mathbf{R}}  - \hat{\mathbf{S}} \right) \hat{\mathbf{b}}, \label{eq:matrixeq2}
\end{eqnarray}
and solve for 
\begin{eqnarray}
 \det \left(  \hat{\mathbf{T}} \hat{\mathbf{R}}  - \hat{\mathbf{S}} \right)/\det \left(\hat{\mathbf{R}} -  \hat{\mathbf{T}} \hat{\mathbf{S}} \right)  =0.  \label{eq:determinant}
\end{eqnarray}
Equation \ref{eq:matrixeq2} represents an infinite matrix system of equations, but it can be truncated to just a few terms by considering the relevant modes. A first order approximation is to consider only the exchange of incident/reflected waves that are separated by just one mode. This results in $4\times4$ matrices, with modes $k-1$ and $k$ involved, and already reproduces some of the main features seen in the full electrodynamic simulations of the gratings. On the other hand, a second order calculation includes waves that are connected by up to two modes ($k-2$, $k-1$, $k$ and $k+1$), resulting in $8\times8$ matrices and giving more accurate results. Finally, a third order calculation - exchange of waves separated by three modes, from $k-3$ to $k+2$, $12\times12$ matrices - produces results that, as we will see, show an almost perfect agreement with simulations.

\begin{figure*}[p]
\begin{center}
\includegraphics [angle=0, width=\textwidth]{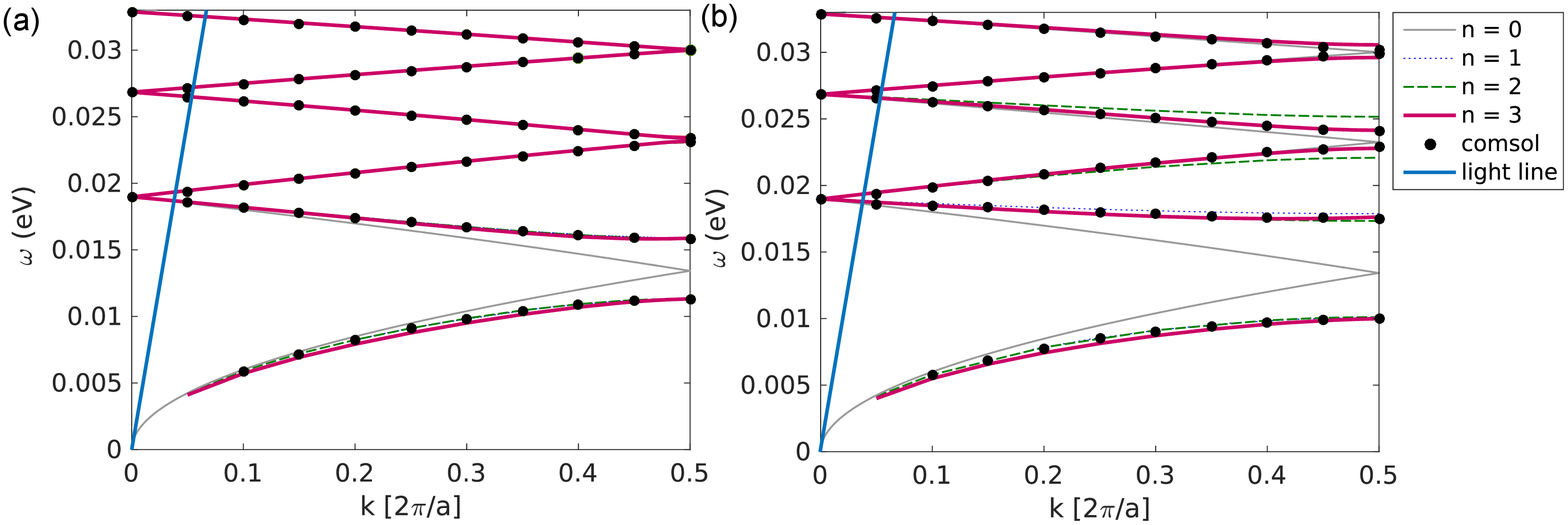}
\caption[]{Dispersion relation in the first Brillouin zone for graphene
conductivity gratings at $\mu = 0.1$ eV, for (a) weak modulation ($w_0=1.5$) and (b) strong modulation ($w_0=2.5$).  The graphene is placed on a substrate ($\epsilon_1=3$, $\epsilon_2=1$) and in both cases the period is $a=2\pi\gamma$, with $\gamma=4\cdot10^{-7}$ m. We start with a graphene layer at $u_0=1$ in the slab frame. The thin grey line represents the plasmon dispersion relation of a homogeneously biased graphene layer (Eq. \ref{eq:graphenesp}), and the dotted blue, dashed green and thick purple lines correspond to the solution for the modes obtained with Eq. \ref{eq:determinant} and including one, two and three modes in the calculation, respectively. The black dots show full electrodynamic simulation results and the dark blue line represents the light line. }\label{fig3}
\end{center}
\end{figure*}

\section{Results}
\label{sec:results}

In this section we apply our formalism to derive the band structure of two instances of plasmonic gratings: graphene conductivity gratings and metal gratings.  

\subsection{Graphene with modulated conductivity}

Compared to other commonly employed plasmonic materials such as nobel metals, graphene presents two main advantages \cite{Vafek2006,Hanson2008,Jablan2009,Koppens2011,Nikitin2011,Fei2011,Fei2012,Chen2012}. First, its characteristic linear dispersion of electrons close to the Dirac point together with its low carrier density implies that the chemical potential of graphene can be dynamically tuned in a broad band range by, e.g., electrostatic gating. This provides biased graphene with the potential of supporting surface plasmons of tuneable properties. In addition, owing to the 2D character of this material, graphene plasmons are extremely confined to the sheet, that is, their wavelength is much smaller than the free space wavelength of radiation, by a factor much larger than the typical factors observed in noble metals. Here we consider a graphene monolayer subject to a spatially periodic bias that results in a periodic modulation of the conductivity of the sheet along one direction \cite{Huidobro2016,Huidobro2016a}. Such a conductivity grating provides the missing momentum for a free space photon at THz frequencies to couple into the surface plasmons in graphene, with wavelengths of a few $\mu$m. Alternative coupling methods into surface plasmons include relief corrugations or periodically patterned substrates \cite{Zhan2012,Slipchenko2013,Stauber2014a,Miao2015,Zhao2015}, as well as patterned graphene structures \cite{Ju2011,Nikitin2012b,Nikitin2012,Thongrattanasiri2012a,Fang2013,Fang2014}.

Within our transformation optics formalism, a graphene conductivity grating can be treated by considering the limit of an infinitesimally thin slab. A graphene layer of conductivity $\sigma_g(\omega)$ can be equivalently represented as a thin slab of material of thickness $d$ and permittivity,
 \begin{equation}
	 \epsilon(\omega)=  1+i  \frac{\sigma_g(\omega)}{\omega\epsilon_0 d}, \label{eq:epsilongraphene}
\end{equation}
Such a thin slab maps to another thin slab of modulated thickness through Eq. \ref{eq:transformation}, as sketched in Fig. 1(a). The slab in the grating frame equivalently represents a graphene layer with periodically varying conductivity \cite{Huidobro2016}. Since conformal transformations preserve the material properties in the plane of incidence, $\epsilon(\omega)$ is the same in both frames, and, according to Eq. \ref{eq:epsilongraphene}, $\sigma_g(\omega)$ varies periodically following the variations of the material thickness. This allows us to apply the formalism presented in Section \ref{sec:methodology} to the case of a graphene conductivity grating. 

The most important result stemming from our transformation optics framework is that the modes of the graphene conductivity grating are linked to those of a simple layer of homogeneously biased graphene. Within the quasi static limit, the dispersion relation of the graphene plasmons propagating along a homogeneous graphene sheet is given by, 
\begin{equation} 
	\epsilon_1+\epsilon_2+i\frac{\sigma_g(\omega)}{\epsilon_0\omega}|k|=0\,, \label{eq:graphenesp}
\end{equation}
with $k$ the component wavevector parallel to the grating.
The conductivity of graphene, $\sigma_g(\omega)$, is taken from the RPA \cite{wunsch2006}. The equation above also represents the dispersion relation of the surface plasmons in the graphene conductivity grating, at least close to the zone centre. 

Figure \ref{fig3} presents the band structures of two graphene conductivity gratings, one weakly modulated, $w_0=1.5$ (left panel), and one strongly modulated, $w_0=2.5$ (right panel). In both cases the grating period is $a=2\pi \gamma$, with $\gamma =  4\cdot10^{-7}$ m, the chemical potential is $\mu=0.1$ eV and the scattering losses are assumed to be low ($\tau=10$ ps). In the figure, the black dots represent numerical results for the eigenfrequencies of the full electrodynamics of the problem. The dispersion relation of plasmons in a homogenous graphene sheet, Eq. \ref{eq:graphenesp} folded in the first Brillouin zone, is plotted  as a thin grey line. It is clear that the dispersion in the slab frame agrees perfectly with the numerically calculated modes of the grating close to the zone centre. The translational invariance of the graphene in the slab frame implies the absence of a band gap at $k=0$, independently of how strongly modulated the grating is, so close to the zone centre the modes of both gratings occur at the same energy. It should be noted that the absence of band gaps at $k=0$ is exact only up to magnetic effects, which give rise to a small band gap that is much smaller than the broadening of the bands due to radiation damping. On the other hand, as the parallel momentum increases away from the zone centre, the exact result starts to deviate from Eq. \ref{eq:graphenesp}. The reason for this is Bragg scattering \cite{Barnes1996}. In particular, for the weakly modulated case, a band gap opens at the zone edge for the lowest band. This is to be expected, and is due to the already discussed discontinuities of the waves at the branch cuts. The effect is strongest for the first band since a weakly modulated grating can be very well approximated to a sinusoidal shape, and hence the first Fourier component is much larger than the rest. Indeed, a very small band gap also opens in the second band. For the strongly modulated grating, band gaps at the zone edge are visible for the three lowest bands. The reason for this is that a stronger modulation has a few more Fourier components than the first one, which provide the momentum needed to open those band gaps. 

In order to analytically calculate the band structure throughout the whole of the Brillouin zone, we apply our method to include the mixing of waves through the branch cuts. We solve Eq. \ref{eq:determinant}, where the scattering matrix, $\mathbf{\hat{T}}$, is obtained by matching the fields in the slab frame, as shown in the SM. For the sake of simplicity in our calculation we assume lossless graphene [$\text{Re}(\sigma_g)=0$], and we have,
\begin{eqnarray}
T_{k}^{++} & = & \frac{2\epsilon_{2}}{\epsilon_{2}+\epsilon_{1}+i\sigma_g(\omega)|k|/(\omega\epsilon_{0})}, \\
T_{k}^{+-} & = & \frac{\epsilon_{2}-\epsilon_{1}-i\sigma_g(\omega)|k|/(\omega\epsilon_{0})}{\epsilon_{2}+\epsilon_{1}+i\sigma_g(\omega)|k|/(\omega\epsilon_{0})}e^{+2|k|u_0},\\
T_{k}^{-+} & = & \frac{\epsilon_{1}-\epsilon_{2}-i\sigma_g(\omega)|k|/(\omega\epsilon_{0})}{\epsilon_{2}+\epsilon_{1}+i\sigma_g(\omega)|k|/(\omega\epsilon_{0})}e^{-2|k|u_0},\\
T_{k}^{--} & = & \frac{2\epsilon_{1}}{\epsilon_{2}+\epsilon_{1}+i\sigma_g(\omega)|k|/(\omega\epsilon_{0})}.
\end{eqnarray}
The dispersion relation obtained using this analytical method is plotted in Fig. \ref{fig3} for three different number of modes included in the calculation: $n=1$, dotted blue line, $n=2$, dashed green line and $n=3$, thick purple line. For the weakly modulated grating, taking only only one mode into account nicely reproduces the first band gap. Introducing a second mode accounts for the contribution that yields to the second band gap opening, and three modes give a very similar result. The results for the grating with strong modulation are similar but the effects are more clearly seen since the first three bands show a band gap. Including the first mode reproduces almost completely the first band gap, and including the second mode brings the results closer to the simulations. Next, introducing this second mode reproduces the second band gap, but it overestimates it. Finally, working with three modes narrows the second band gap, and gives rise to a third band gap. Again, this third band gap is slightly overestimated compared to the simulation results, but that could be solved by introducing one more mode in the calculation.

\begin{figure*}[p]
\begin{center}
\includegraphics [angle=0, width=\textwidth]{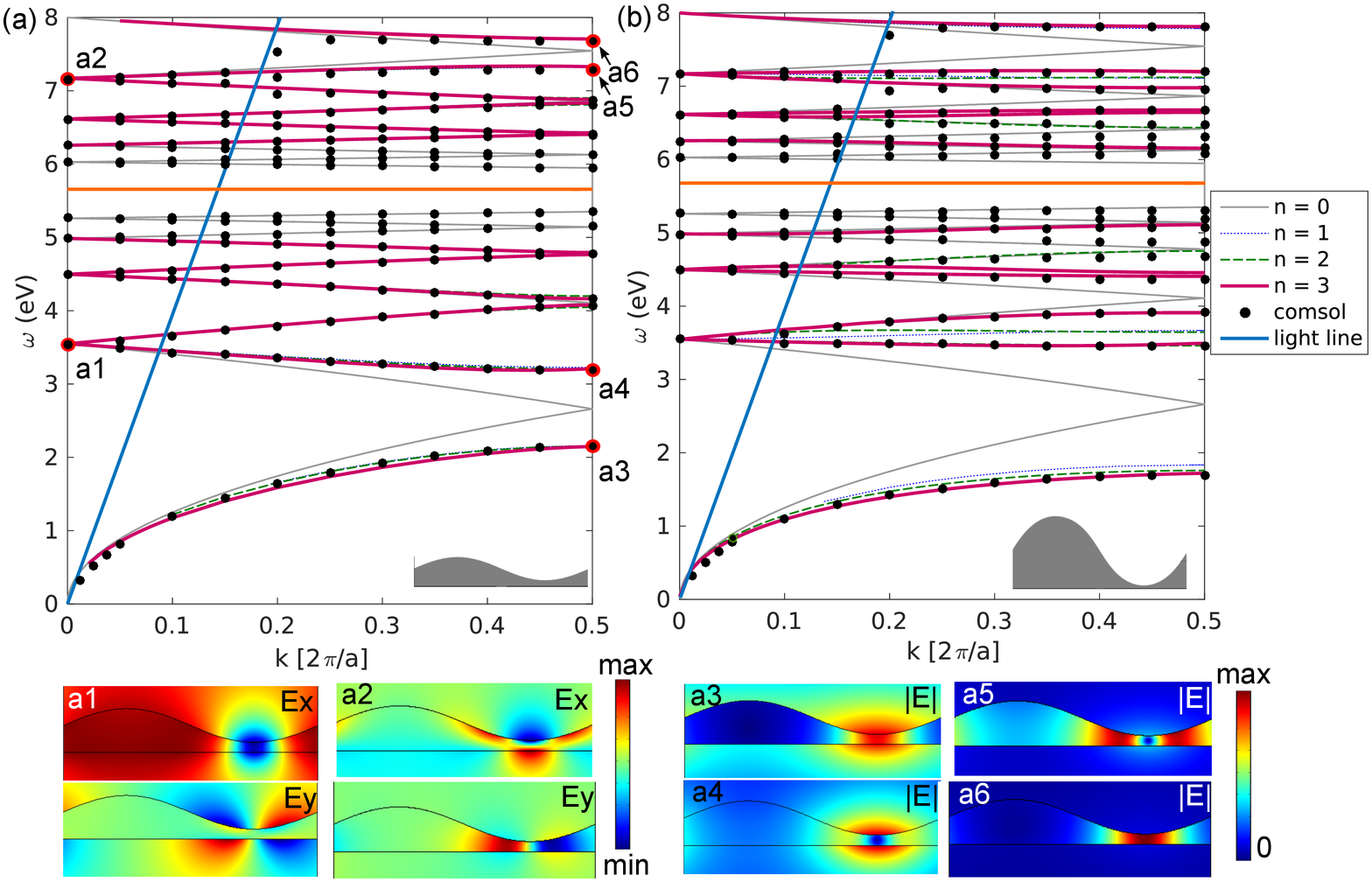}
\caption[]{Dispersion relation for silver gratings of two different modulation strengths: $w_0=1.5$ (a) and $w_0=2.5$ (b). The period of both gratings is $a=2\pi\gamma$, with $\gamma=5\cdot10^{-9}$ m, and $\epsilon_1=\epsilon_2=1$. We start with a metal film of thickness $d=0.5$ at $u_0=1$ in the slab frame. Modes below and above the surface plasmon frequency, $\omega_p/\sqrt{2}$, marked with an orange line, are shown. Simulation results are plotted as black dots, while the lines correspond to our analytical theory as in Fig. \ref{fig3}. Note that, especially for the weakly modulated grating, the results for $n=1$ (dotted blue line), in the region of the first band gap, almost overlap with the results for $n=3$ (thick purple line). Similarly, the results for $n=2$ (dashed green line) for the first and second band gaps are almost the same as the results for $n=3$. Note also that the modes in the vicinity of the surface plasmon frequency (orange line) have been removed for the sake of clarity. Lower inset panels: electric field patterns in the weakly modulated grating at the selected points at $k=0$ and $k=\pi/a$ as indicated in panel (a).}\label{fig4}
\end{center}
\end{figure*}

\subsection{Metal grating}

We now consider the modes of subwavelength metal gratings. Plasmonic gratings provide a simple scheme within state of the art nanofabrication to couple radiation into modes that are confined in a scale much smaller than the wavelength. Similarly to the graphene conductivity gratings that we have discussed, a periodic pattern in a metal film enables the coupling of a plane wave to higher order modes diffracted by the grating, and the associated energy concentration has been used for applications in biosensing as well as solar cells \cite{Wen2014,Munday2011,Dhawan2011,Bog2012,Kuo2012}. In addition, plasmonic gratings provide a simple design for asymmetric reflectors (or absorbers) in the optical regime due to biansotropy \cite{Kraft2016}.

Following our transformation optics framework, the modes of the metal grating can be mapped to those of a slab of thickness $d$, which in the quasistatic limit follow the well known dispersion relation, 
\begin{equation}
	e^{|k|d} =\pm \frac{\epsilon_m(\omega)-1}{\epsilon_m(\omega)+1} \label{eq:metalsp}
\end{equation}
Figure \ref{fig4} presents the band structures of a weakly (left panel) and a strongly (right panel) modulated silver grating. The permittivity of silver is described with a Drude model, 
\begin{equation}
	\epsilon_m(\omega) = 1-\frac{\omega_p^2}{\omega(\omega+i\gamma)},
\end{equation} 
with plasma frequency $\omega_p=8$ eV. Again, for our analytical calculations we take a lossless metal. The dispersion relation of the silver grating in the first Brillouin zone is shown for frequencies below $\omega_p$. The solution to Eq. \ref{eq:metalsp}, that is, the dispersion of the flat slab, is plotted as a thin grey line. In the slab frame, the $k$ vector increases greatly as the frequency approaches the surface plasmon resonance, $\omega_{sp}=\omega_p/\sqrt{2}$. Folding the surface plasmon dispersion relation in the first Brillouin zone hence results in the modes being closer together as the frequency approaches $\omega_{sp}$ both from below and above. In fact, the surface plasmon frequency separates modes of different symmetries: bonding modes below $\omega_{sp}$, antibonding modes above  $\omega_{sp}$ [see inset panels (a1) and (a2) in Fig. \ref{fig4}]. Close to the zone centre, Eq. \ref{eq:metalsp} shows an almost perfect agreement with the results obtained from full electrodynamics simulations (black dots) above and below $\omega_{sp}$. On the other hand, and as we have explained before, the dispersion in the slab frame fails to reproduce the band gap openings at the zone edge. 

In order to derive the full band structure with our formalism we first write the scattering matrices for this case (see SM), which read as 
\begin{eqnarray}
	T^{++}_k	&=& e^{2|k|u_0} \frac{1-e^{2\alpha}}{e^{2|k|d}-e^{2\alpha}}, \\
	T^{+-}_k	&=& -e^{-2|k|(u_0+d)} \frac{e^{2|k|d}-1}{e^{2|k|d}-e^{2\alpha}}, \\ 
	T^{-+}_k	&=& -e^{-2|k|u_0} \frac{e^{2|k|d}-1}{e^{2|k|d}-e^{2\alpha}},  \\
	T^{--}_k	&=& e^{2|k|d} \frac{1-e^{2\alpha}}{e^{2|k|d}-e^{2\alpha}}, 
\end{eqnarray}
The dispersion relation obtained by solving Eq. \ref{eq:determinant} is plotted in Fig. \ref{fig4}. For the case of the weak grating, the numerical results show how the lowest energy band below $\omega_{sp}$ , i.e. the lowest bonding mode, and the highest energy band below $\omega_p$, i.e. the highest bonding mode, present the largest band gaps at $k=\pi/a$. Taking into account one mode (dotted blue line) gives the the lowest and highest energy bands, in good agreement with numerics and reproducing both band gap openings. Including two modes in the calculation (dashed green line) slightly improves the result for the lowest and highest bands, and also reproduces the small band gap openings for the second lowest and second highest bands, in an excellent agreement with simulations. Next, the calculation with 3 modes also reproduces the third band below $\omega_{sp}$ and third band below $\omega_{p}$. On the other hand, for the grating with strongest modulation, the interaction effects between bands are more pronounced and the band gaps at $k=\pi/a$ are much larger. In this case, including only one mode in the formalism (dotted blue line), fails to accurately predict the size of the lowest and highest band gaps and shows some deviation from the numerical results for the bands. Similar features occur for the analytical results with 2 modes (dashed green line). The results for $n=3$ (thick purple line) show an almost perfect agreement with simulations for the lowest and highest band gap openings, while the next order ones can be improved by considering one more mode in the formalism.   

Finally, we discuss here how our analytical predictions show small deviations at certain frequencies from the bands obtained with full electrodynamics. As shown in Fig. 4, some of the bands above the surface plasmon frequency bend when approaching the light line and avoid crossing it, whereas our quasistatic solution predicts a direct crossing of the light line. This effect is more pronounced for the highest frequency band, and is reminiscent of the behaviour of the antibonding mode supported by a simple metal slab. The retarded dispersion relation for the antibonding mode approaches the quasistatic solution, $\omega_{sp}$ at high $k$ vectors, and grows to higher frequencies for smaller $k$ vectors. As the light line is approached, the dispersion bends towards it and tends towards zero frequency. In fact, this perturbation of the bands was almost absent for the graphene conductivity grating discussed above since in the limit of an infinitely thin slab only the bonding mode survives.

\section{Conclusions}
\label{sec:conc}
In this paper, we studied plasmonic gratings that inherit their spectral properties from a simple translationally invariant thin slab. The grating and the slab are related to each other through a conformal map, such that the plasmonic response of the grating is the same as that for the slab. This statement is exact only at normal incidence, where we have shown how the incident wave does not couple to the grating, and the dispersion relation is continuous regardless the grating strength. On the other hand,  out of normal incidence the band structure displays band gap openings where the dispersion relation is no longer continuous. This is an effect of the distortion of the geometry when going from the slab to the grating, which is reflected by branch points in the conformal transformation. By appropriately taking into account these singularities, which result in incident (transmitted) waves transforming into transmitted (reflected) waves, we accurately calculate the dispersion of the plasmon modes supported by the grating in whole frequency range and for the whole Brillouin zone. 

We have studied gratings formed by a thin metal film with one periodically curved surface and to gratings obtained by doping a graphene layer with a spatially periodic pattern. In the case of graphene gratings, our analytical treatment results in dispersion relations that are in perfect agreement with those obtained from full electrodynamics simulations. This is due to the 2D character of the graphene, which results in very confined surface plasmons that can be very well described under the quasistatic approximation. On the other hand, for the case of metal gratings, the band structure shows some effects of retardation close to the light line for the antibonding modes. While this cannot be captured by our quasistatic treatment, we have shown that our formalism describes almost perfectly the physics behind the band structure. In fact, a very good agreement with full electrodynamics calculations is obtained for very subwavelength metal gratings. Our analytical treatment allows for an easy optimization of the grating parameters for a given application.

\section*{acknowledgements}
We are grateful for fruitful discussions with Yu Luo. This work was supported by the Leverhulme Trust, the EPSRC (Grant No. EP/L024926/1), and the Gordon and Betty Moore Foundation. P.A.H. acknowledges funding from a Marie Sklodowska-Curie Fellowship.

\bibliography{plasmonic_gratings}

\end{document}